\documentstyle[12pt]{ioplppt}                

\begin{document}

\jl{1}

\title{A class of integrable lattices and KP hierarchy}

\author{A K Svinin\ftnote{1}{E-mail address : svinin@icc.ru}}

\address{Institute of System Dynamics and Control Theory,
Siberian Branch of Russian Academy of Sciences,
P.O. Box 1233, 664033 Irkutsk, Russia}

\begin{abstract}
We introduce a class of integrable $l$-field first-order lattices
together with corresponding Lax equations. These lattices may be
represented as
consistency condition for auxiliary linear systems defined on sequences
of formal dressing operators. This construction provides simple
way to build lattice Miura transformations between one-field lattice
and $l$-field ($l\ge 2$) ones. We show that the lattices
pertained to above class is in some sense compatible with KP
flows and define the chains of constrained KP Lax operators.
\end{abstract}

\maketitle

\section{Introduction}

In recent times, the Kadomtsev--Petviashvili (KP) hierarchy and its
possible reductions has drawn much attention because of vast variety
of applications to different branches of physics. Recall that in
Sato framework the KP hierarchy is given by Lax equation
\begin{equation}
\partial_p{\cal Q} = [({\cal Q}^p)_{+}, {\cal Q}]
\label{lax0}
\end{equation}
on first-order pseudo-differential operator ($\Psi$DO) ${\cal Q} = \partial +
\sum_{k=1}^{\infty}U_k(\underline{t})\partial^{-k}$. Equivalent form
of Lax equation (\ref{lax0}) is Sato--Wilson equation
\begin{equation}
\partial_p\hat{w} = - (\hat{w}\partial^p\hat{w}^{-1})_{-}\hat{w} =
(\hat{w}\partial^p\hat{w}^{-1})_{+}\hat{w} - \hat{w}\partial^p
\label{sato}
\end{equation}
on a formal dressing operator $\hat{w}$ being defined through
${\cal Q} = \hat{w}\partial\hat{w}^{-1}$. Another objects associated
with ${\cal Q}$ are formal Backer--Akhiezer wave function $\psi$ and its
conjugate $\psi^{*}$
defined by $\psi(\underline{t}, z) = \hat{w}\exp(\xi(\underline{t}, z))$ and
$\psi^{*}(\underline{t}, z) = \hat{w}^{*-1}\exp(-\xi(\underline{t}, z))$,
respectively,
with $\xi(\underline{t}, z) = \sum_{p=1}^{\infty}t_pz^p$.

The very important observation is that the KP wave functions satisfy the
bilinear identity \cite{date}
\begin{equation}
{\rm res}_z\psi(\underline{t}, z)\psi^{*}(\underline{t}^{'}, z) = 0
\label{identity}
\end{equation}
providing description of KP hierarchy in terms of Hirota's bilinear
equations on $\tau$-function. After introducing $\tau(\underline{t})$
through
\[
\psi(\underline{t}, z) =
\frac{\tau(\underline{t} - [z^{-1}])}
{\tau(\underline{t})}\exp(\xi(\underline{t}, z)),\;\;
\psi^{*}(\underline{t}, z) =
\frac{\tau(\underline{t} + [z^{-1}])}
{\tau(\underline{t})}\exp(-\xi(\underline{t}, z))
\]
the identity (\ref{identity}) becomes
\[
{\rm res}_z\tau(\underline{t} - [z^{-1}])\tau(\underline{t}^{'} + [z^{-1}])
\exp(\xi(\underline{t}-\underline{t}^{'}, z)) = 0
\]
with $[z^{-1}] = (1/z, 1/(2z^2), 1/(3z^3),...)$.

In present article we suggest construction
which provides relationship between some class of multi-field lattices and chains
of constrained KP Lax operators. Recently constrained KP (cKP) hierarchies
gained a lot of interest because of its strong relationship
with multi-matrix models in non-perturbative string theory \cite{bonora},
\cite{aratyn1}. As is known cKP hierarchies contain a number
of interesting, from the physical point of view, integrable nonlinear
evolution equations whose applications range from hydrodynamics to
modern theories in high-energy elementary particle physics. Important
property of cKP hierarchies is the existence of discrete
symmetry structure defined by successive Darboux -- B\"acklund
transformations of suitable Lax operators \cite{aratyn1}.
The integrable lattices play key r\^ole in this situation
and serve as a source of discrete symmetries for evolution
type systems \cite{levi}, \cite{leznov}, \cite{aratyn2} ``gluing"
a copies of the equations of cKP.
In particular it is well known relationship between Toda and Volterra
lattices and some class of cKP hierarchies.

We exhibit in this paper a class of integrable
$l$-field lattices.
They are defined as consistency condition
of some auxiliary linear systems. In fact these auxiliary systems
define compatible pairs of shifts $s_1$ and $s_2$ on sequences of dressing
operators $\{\hat{w}_i, \in {\bf Z}\}$.
This construction naturally provides
building of infinite sets of constrained Lax operators  ``glued" together
by compatible pair of similarity gauge transformations. The important
property of the lattices that they are consistent in a sense with KP
flows given by Lax equation (\ref{lax0}) or Sato--Wilson equation (\ref{sato}).

This paper in many respects was influenced by the
work \cite{dickey} where the system
\[
(\partial + u_i)\hat{w}_i = \hat{w}_{i+1}\partial,
\]
\[
\partial_p\hat{w}_i = - ({\cal Q}_i^p)_{-}\hat{w}_i,\;\;{\rm where}\;\;
{\cal Q}_i = \hat{w}_i\partial^p\hat{w}_i^{-1}
\]
\[
\partial_pu_i = ({\cal Q}_{i+1}^p)(\partial + u_i) -
(\partial + u_i)({\cal Q}_i^p),\; i\in {\bf Z}
\]
was introduced and referred to as modified KP hierarchy. It
was shown there that modified KP hierarchy in fact is
equivalent to discrete KP (1-Toda lattice) hierarchy. Lax operators
${\cal Q}_i$ in this case are unconstrained and only connected with
each other by similarity gauge transformations. Our construction
provides building of sequences of constrained KP copies.
As we have mentioned above this is achieved by imposing on the sequences
$\{\hat{w}_i,\; i\in{\bf Z}\}$ two compatible auxiliary linear
constraints. A compatibility of the latter is guaranteed
by equations of integrable lattices. Another important inspiration
of this work is Krichever's rational reductions of the KP hierarchy
\cite{krichever} (or constrained KP, as these commonly called).

A scheme of the paper is as follows. In section 2 we present integrable chains.
In section 3 we show how one can construct Miura lattice transformations and
provide the reader by some examples.
Also we write down Miura transformations in explicit form for two-field
lattices. In section 4 we show compatibility of the lattices with KP
flows.

\section{Integrable lattices}

Main objective of this section is to define a class of an integrable lattices.
The term ``integrable" in our context means only that given lattice
admits Lax representation. We will consider first-order
differential-difference systems on finite number of fields $a_k(i, x)$ being
functions of discrete
variable $i\in{\bf Z}$ and continuous variable $x\in{\bf R}^1$.

Before proceeding further let us give some remarks concerning notations which
will be used in the following. The unknown
functions (fields) depend on spatial variable $x\in{\bf R}^1$ and some
evolution parameters $t_p$. We use short-hand notation $\underline{t}$
to denote infinite collection of independent variables
$(x, t_2, t_3,...)$. The symbols $\partial$ and $\partial_p$
stand for derivation with respect to $x$ and $t_p$, respectively.
Let $A$ be a $\Psi$DO $A = \sum_{i=-\infty}^{N}a_i(x)\partial^{i}$, of order $N$.
The subscripts
$+$ and $-$ mean, respectively, differential and integral part of $A$.
We write
$\partial_p A$ to denote derivative of $\Psi$DO $A$ with respect to
$t_p$ (not a product).

For an arbitrary pair of integers $n\in{\bf N}$  and $m \le n-1$ we
define infinite collection of first-order differential operators
\begin{equation}
H_i = \partial - \sum_{k=1}^na_0(i+k-1, x),\;\;
i\in{\bf Z},
\label{H_i}
\end{equation}
and $\Psi$DO's
\[
G_i = \partial + \sum_{k=1}^{|m|}a_0(i-k, x) +
\sum_{k=1}^{l-1}a_k(i+m, x)H_{i-kn}^{-1}...H_{i-2n}^{-1}H_{i-n}^{-1},
\]
for $m\le -1$ and
\[
G_i = \partial - \sum_{k=1}^{m}a_0(i+k-1, x) +
\sum_{k=1}^{l-1}a_k(i+m, x)H_{i-kn}^{-1}...H_{i-2n}^{-1}H_{i-n}^{-1},
\]
for $m = 0,..., n-1$.
Notice that  definition of $H_i$'s and $G_i$'s involves a finite number of
fields
$\{a_0(i, x),  a_1(i, x),..., a_{l-1}(i, x)\}$.

Let us define following auxiliary equations on infinite collection
of dressing operators $\{\hat{w}_i,\; i\in{\bf Z}\}$:
\begin{equation}
G_i\hat{w}_i = \hat{w}_{i+m}\partial,\;\;\;
H_i\hat{w}_i = \hat{w}_{i+n}\partial
\label{aux1}
\end{equation}
Obviously the latter can be rewritten in terms of BA functions as
\begin{equation}
G_i\psi_i = z\psi_{i+m},\;\;\;
H_i\psi_i = z\psi_{i+n}.
\label{aux2}
\end{equation}
The linear system (\ref{aux2}) (or (\ref{aux1})) is overdetermined
but one can show that the compatibility conditions of (\ref{aux2})
are well-determined system of equations for the fields $a_k(i, x)$.

Formally,
consistency condition of (\ref{aux2}) is given by
\begin{equation}
G_{i+n}H_i = H_{i+m}G_i
\label{compat}
\end{equation}
or equivalently as $H_{i+m}^{-1}G_{i+n} = G_iH_i^{-1}$.
The relation (\ref{compat}) is not convenient for further calculations.
The technical observation which is helpful in this situation
is that (\ref{aux2}) can be rewritten in terms of
$({\bf L}, {\bf A})$-pair
\[
{\bf L}(\psi_i) = z\psi_i,\;\;\;  \psi_i^{\prime} = {\bf A}(\psi_i)
\]
with ${\bf L}$ and ${\bf A}$ being
difference operators acting on the space of sequences of BA functions
$\{\psi_i,\;\; i\in{\bf Z}\}$ as
\begin{equation}
\begin{array}{l}
\displaystyle
{\bf L}(\psi_i) = z\psi_{i+n-m} + \left(\sum_{s=1}^{n-m}a_0(i+s-1)\right)\psi_{i-m} +
\sum_{j=1}^{l-1}\frac{1}{z^j}a_j(i)\psi_{i-m-jn}, \\[0.4cm]
\displaystyle
{\bf A}(\psi_i) = z\psi_{i+n} + \left(\sum_{s=1}^{n}a_0(i+s-1)\right)\psi_{i}.
\end{array}
\label{la0}
\end{equation}
Then consistency condition of (\ref{aux2}) are expressed in a form
of the Lax equation
\begin{equation}
{\bf L}^{\prime} = [{\bf A}, {\bf L}] =
{\bf A}{\bf L} - {\bf L}{\bf A}.
\label{lax}
\end{equation}
One can show that the latter holds if the functions
$a_k(i, x)$ satisfy $l$-field lattice
\begin{equation}
\begin{array}{l}
\displaystyle
\sum_{s=1}^{n-m}a_{0}^{\prime}(i+s-1)=
\sum_{s=1}^{n-m}a_{0}(i+s-1) \\[0.4cm]
\displaystyle
\times\left(\sum_{s=1}^{n}a_{0}(i+s-1)-
\sum_{s=1}^{n}a_{0}(i+s-m-1)\right) \\[0.4cm]
\displaystyle
+ a_1(i+n) - a_1(i), \\[0.4cm]
\displaystyle
a_{k}^{\prime}(i)=
a_{k}(i)\left(\sum_{s=1}^{n}a_{0}(i+s-1)-
\sum_{s=1}^{n}a_{0}(i+s-m-kn-1)\right) \\[0.4cm]
\displaystyle
+ a_{k+1}(i+n) - a_{k+1}(i),\;\;\;
k = 1,..., l-1.
\end{array}
\label{1}
\end{equation}
Here it is understood that $a_l(i, x)=0$.

{\bf Example 2.1} Consider the case $n=1, m=-1, l=2$. The system
(\ref{1}) becomes
\[
a_0^{\prime}(i) + a_0^{\prime}(i+1)
\]
\[
= (a_0(i) + a_0(i+1))(a_0(i) - a_0(i+1)) + a_1(i+1) - a_1(i),
\]
\[
a_1^{\prime}(i) = 0.
\]
So, actually we have in this case one-field lattice\footnote{For one-field
lattices we use notation $a_0(i)=r_i$}
\begin{equation}
r_i^{\prime} + r_{i+1}^{\prime} = r_i^2 - r_{i+1}^2 + \nu_i,
\label{10}
\end{equation}
with $\nu_i = a_1(i+1) - a_1(i)$ being some constants. As is known the
lattice (\ref{10}) describes elementary Darboux transformation for
Schr\"odinger operator $L = \partial^2 - q(x)$.
An interesting property of the lattice (\ref{10}) is that it
reduces to Painlev\'e transcedents $P_4$ and $P_5$ due to imposing
periodicity conditions
\[
r_{i+N} = r_i,\;\; \nu_{i+N} = \nu_i
\]
for $N=3$ and $N=4$, respectively \cite{adler}.

{\bf Example 2.2}  In the case $m=0,\: n=1,\: l\ge 2$ we obtain well known
generalized Toda systems
\begin{equation}
\begin{array}{l}
\displaystyle
a_{0}^{\prime}(i) = a_1(i+1) - a_1(i), \\[0.4cm]
\displaystyle
a_{k}^{\prime}(i) = a_{k}(i)\left(a_0(i) - a_0(i-k)\right)  \\[0.4cm]
+ a_{k+1}(i+1)-a_{k+1}(i),\;\;
k = 1,..., l-1.
\end{array}
\label{toda}
\end{equation}
In particular if $l=2$ we obtain ordinary Toda lattice in
polynomial form
\begin{equation}
\begin{array}{l}
\displaystyle
a_{0}^{\prime}(i) = a_1(i+1) - a_1(i), \\[0.4cm]
\displaystyle
a_{1}^{\prime}(i) = a_{1}(i)\left(a_0(i) - a_0(i-1)\right).
\end{array}
\label{TODA}
\end{equation}
Defining $u_i$ by relation $a_0(i) = -u_i^{\prime}$ and
$a_1(i) = \exp(u_{i-1}-u_{i})$ we arrive at more familiar exponential
form of the Toda lattice $u_i^{\prime\prime} = e^{u_{i-1}-u_i} - e^{u_i-u_{i+1}}.$

{\bf Example 2.3} Let $m = n-1,\: l = 1,\: n\ge 2$. This choice corresponds
to Bogoyavlenskii lattices
\cite{bogoyavlenskii}
\begin{equation}
r_{i}^{\prime} = r_{i}\left(\sum_{k=1}^{n-1}r_{i+k} -
\sum_{k=1}^{n-1}r_{i-k}\right).
\label{volterra}
\end{equation}

\section{Miura lattice transformations}

Representation of the chains (\ref{1}) as consistency
condition of an auxiliary linear equations (\ref{aux2}) allows us to
construct in simple and algorithmical way Miura mapping which connect
solutions of one-field lattices with solutions of corresponding
$l$-field ($l\ge 2$) ones.

Firstly notice that $F_i = G_{i+(l-1)n}H_{i+(l-2)n}...H_{i+n}H_i$
is $l$-order differential operator. As consequence of (\ref{aux2})
we obtain
\begin{equation}
F_i\psi_i = z^l\psi_{i+(l-1)n+m},\;\;\;
H_i\psi_i = z\psi_{i+n}.
\label{aux5}
\end{equation}
Here it is important to notice that (\ref{aux5}) in fact is equivalent
to (\ref{aux2}). Indeed one can express $G_i$ as $G_i =
F_{i-(l-1)n}H_{i-(l-1)n}^{-1}...H_{i-2n}^{-1}H_{i-n}^{-1}$ and obtain
(\ref{aux2}) as consequence of (\ref{aux5}).

Define two integers
\begin{equation}
\overline{n} = ln,\;\;\;
\overline{m} = (l-1)n+m.
\label{para}
\end{equation}
It is evident that $\overline{n}\ge 2$ and  $\overline{m}<\overline{n}$.
Let us identify $\psi_i = \overline{\psi}_{li}$,
where $\overline{\psi}_{i}$ is BA functions being determined by
auxiliary linear system
\begin{equation}
\overline{G}_i\overline{\psi}_i = z\overline{\psi}_{i+\overline{m}},\;\;\;
\overline{H}_i\overline{\psi}_i = z\overline{\psi}_{i+\overline{n}}.
\label{aux6}
\end{equation}
with $\overline{H}_i = \partial - \sum_{k=1}^{\overline{n}}r_{i+k-1}$ and
\[
\overline{G}_i = \left\{
\begin{array}{ll}
\displaystyle
\partial + \sum_{k=1}^{|\overline{m}|}r_{i-k},&{\rm for}\;\;\overline{m}\le -1, \\[0.4cm]
\displaystyle
\partial - \sum_{k=1}^{\overline{m}}r_{i+k-1},&{\rm for}\;\;\overline{m}\ge 1.
\end{array}
\right.
\]
Consistency condition of auxiliary equations (\ref{aux6}) is equivalent
to one-field lattice
\begin{equation}
\sum_{s=1}^{\overline{n}-\overline{m}}r_{i+s-1}^{\prime} =
\sum_{s=1}^{\overline{n}-\overline{m}}r_{i+s-1}\left(
\sum_{s=1}^{\overline{n}}r_{i+s-1} -
\sum_{s=1}^{\overline{n}}r_{i+s-\overline{m}-1}
\right).
\label{chain1}
\end{equation}
As consequence of auxiliary equations (\ref{aux6}) we have
\begin{equation}
\overline{F}_i\overline{\psi}_i = z^l\overline{\psi}_{i+l\overline{m}},\;\;
\overline{H}_i\overline{\psi}_i = z\overline{\psi}_{i+\overline{n}},
\label{aux7}
\end{equation}
with $\overline{F}_i = \overline{G}_{i+(l-1)\overline{m}}...\overline{G}_{i+\overline{m}}\overline{G}_{i}.$
Comparing (\ref{aux7}) with (\ref{aux5}) we arrive at the following identification
\begin{equation}
F_i = \overline{F}_{li},\;\;
H_i = \overline{H}_{li}.
\label{identify}
\end{equation}

Let us exhibit some examples of Miura transformations calculated by using
(\ref{identify}).

{\bf Example 3.1} Take, for example $\overline{n}=2$ and $\overline{m}=1$.
Solving (\ref{para}) gives $n=1$, $m=0$ and $l=2$. In this case
we derive well known relations
\begin{equation}
a_0(i) = r_{2i} + r_{2i+1},\;\;
a_1(i) = r_{2i-1}r_{2i}.
\label{miura0}
\end{equation}
defining a mapping of solutions of the
Volterra lattice
\begin{equation}
r_i^{\prime} = r_i(r_{i+1} - r_{i-1})
\label{v}
\end{equation}
into solutions of the Toda lattice (\ref{TODA}) \cite{miura}.

{\bf Example 3.2} For the system
\begin{equation}
\begin{array}{l}
\displaystyle
a_{0}^{\prime}(i) = a_1(i+1) - a_1(i), \\[0.4cm]
\displaystyle
a_{1}^{\prime}(i) = a_{1}(i)\left(a_0(i) - a_0(i-1)\right) + a_2(i+1) - a_2(i), \\[0.4cm]
a_{2}^{\prime}(i) = a_{2}(i)\left(a_0(i) - a_0(i-2)\right)
\end{array}
\label{TODA1}
\end{equation}
we obtain following Miura transformation
\[
a_0(i) = r_{3i} + r_{3i+1} + r_{3i+2},
\]
\[
a_1(i) = r_{3i-2}r_{3i} + r_{3i-1}r_{3i} + r_{3i-1}r_{3i+1},\;\;
\]
\[
a_2(i) = r_{3i-4}r_{3i-2}r_{3i}
\]
The latter relates (\ref{TODA1}) to Bogoyavlenskii lattice
$r_i^{\prime} = r_i(r_{i+2} + r_{i+1} - r_{i-1} - r_{i-2})$.

>From (\ref{para}) follows that the same one-field lattice is
connected by Miura transformations, generally speaking, with a
number of $l$-field ones.
It is obvious that the number of such lattices is defined by
the fact how many divisors of $\overline{n}$ are among
$l = 2,..., \overline{n}$.

{\bf Example 3.3} Consider Bogoyavlenskii lattice
\begin{equation}
r_i^{\prime} = r_i(r_{i+3} + r_{i+2} + r_{i+1} - r_{i-1} - r_{i-2} - r_{i-3})
\label{volterra1}
\end{equation}
corresponding to the choice $\overline{n} = 4$ and $\overline{m} = 3$
in (\ref{chain1}). Equations (\ref{para}) in this case have two
solutions: $n=2,\; m=1,\; l=2$ and $n=1,\; m=0,\; l=4$.
For the first solution of (\ref{para}) we obtain Miura transformation
\[
a_0(i) = r_{2i} + r_{2i+1},\;\;\; a_1(i) = r_{2i-3}r_{2i}
\]
relating (\ref{volterra1}) to two-field system
\[
a_0^{\prime}(i) = a_0(i)(a_0(i+1) - a_0(i-1)) + a_1(i+2) - a_1(i),
\]
\[
a_1^{\prime}(i) = a_1(i)(a_0(i+1) + a_0(i) - a_0(i-2) - a_0(i-3)).
\]
Second solution of (\ref{para}) corresponds to generalized Toda
lattice  (\ref{toda}) in the case $l=4$, i.e.
\[
\begin{array}{l}
a_{0}^{\prime}(i) = a_1(i+1) - a_1(i), \\[0.4cm]
a_1^{\prime}(i) = a_1(i)\left(a_0(i) - a_0(i-1)\right) + a_{2}(i+1)-a_{2}(i), \\[0.4cm]
a_2^{\prime}(i) = a_2(i)\left(a_0(i) - a_0(i-2)\right) + a_{3}(i+1)-a_{3}(i), \\[0.4cm]
a_3^{\prime}(i) = a_3(i)\left(a_0(i) - a_0(i-3)\right).
\end{array}
\]
Miura transformation in this case is given by
\[
a_0(i) = r_{4i} + r_{4i+1} + r_{4i+2} + r_{4i+3},
\]
\[
a_1(i) = r_{4i-3}r_{4i} + r_{4i-2}r_{4i+1} + r_{4i-1}r_{4i+2} +
r_{4i-2}r_{4i} + r_{4i-1}r_{4i} + r_{4i-1}r_{4i+1},
\]
\[
a_2(i) = r_{4i-6}r_{4i-3}r_{4i} + r_{4i-5}r_{4i-3}r_{4i} +
r_{4i-5}r_{4i-2}r_{4i} + r_{4i-5}r_{4i-2}r_{4i+1},
\]
\[
a_3(i) = r_{4i-9}r_{4i-6}r_{4i-3}r_{4i}.
\]

Let us exhibit results of calculations of Miura transformations
for two-field systems. It can be written in unique form
\[
a_0(i) = r_{2i} + r_{2i+1},\;\;\;
a_1(i) = \sum_{s=1}^{n-m}r_{2i+s-n-m-1}\cdot
\sum_{s=1}^{n-m}r_{2i+s-1}
\]
Notice that the systems corresponding to $m\le -1$ and $n=|m|$ are excluded
from consideration since we have $\overline{m}=0$.
In fact we deal in this situation with one-field lattices
since  $a_1^{\prime}(i) = 0$.

\section{The chains of KP Lax operators}

The relations (\ref{compat}) play key r\^ole for defining Lax
operators ${\cal Q}_i$ connected with each other by compatible
pair of similarity transformations. The subscript $i\in{\bf Z}$
can be interpreted as discrete evolution parameter. The principal
problem naturally raised here is to define equations for
the fields $a_k(i) = a_k(i, \underline{t})$ which guarantee the
compatibility of the mappings with respect to $i$ with $t_p$-flows
given by Lax equation (\ref{lax0}) or equivalently by Sato-Wilson
equation (\ref{sato}).

{\bf Proposition.} {\it By virtue (\ref{compat}), Lax operators are
connected with each other by two invertible compatible gauge
transformations
\begin{equation}
{\cal Q}_{i+m} = G_i{\cal Q}_iG_i^{-1},
\label{similarity1}
\end{equation}
\begin{equation}
{\cal Q}_{i+n} = H_i{\cal Q}_iH_i^{-1}.
\label{similarity2}
\end{equation}
}

{\bf Proof.} By virtue (\ref{compat}), we have
\[
{\cal Q}_{i+m} = \hat{w}_{i+m}\partial\hat{w}_{i+m}^{-1} =
(G_i\hat{w}_{i}\partial^{-1})\partial(\partial\hat{w}_{i}^{-1}G_i^{-1})
\]
\begin{equation}
= G_i\hat{w}_{i}\partial\hat{w}_{i}^{-1}G_i^{-1} = G_i{\cal Q}_iG_i^{-1}.
\label{13}
\end{equation}
The similar calculations are needed to prove (\ref{similarity2}).
The mapping ${\cal Q}_i\rightarrow\tilde{{\cal Q}}_i = {\cal Q}_{i+m}$ we
denote as $s_1$, while $s_2$ stands for transformation
${\cal Q}_i\rightarrow\overline{{\cal Q}}_i = {\cal Q}_{i+n}$. The
compatibility of $s_1$ and $s_2$ also follows from (\ref{compat}).
Indeed, we obtain
\[
{\cal Q}_{i+n+m} = G_{i+n}{\cal Q}_{i+n}G_{i+n}^{-1} =
G_{i+n}H_i{\cal Q}_{i}H_i^{-1}G_{i+n}^{-1}
\]
\[
= H_{i+m}G_{i}{\cal Q}_iG_i^{-1}H_{i+m}^{-1} =
H_{i+m}{\cal Q}_{i+m}H_{i+m}^{-1}.
\]
So we can write $s_1\circ s_2 = s_2\circ s_1$. The inverse maps
$s_1^{-1}$ and $s_2^{-1}$ are well defined by the formulas
${\cal Q}_{i-m} = G_{i-m}^{-1}{\cal Q}_iG_{i-m}$ and
${\cal Q}_{i-n} = H_{i-n}^{-1}{\cal Q}_iH_{i-n}$.   $\Box$

Let $p$ and $q$ are relatively prime integers such that $pn = qm$.
Without loss of generality we can thought that $p\ge 0$. It is
obvious that relation $s_1^q = s_2^p$ holds. Indeed the left-
and right-hand side of this relation correspond to the same mapping
${\cal Q}_i\rightarrow {\cal Q}_{i+pn} = {\cal Q}_{i+qm}$. Let us
summarize the statements above in the following theorem.

{\bf Theorem.} {\it Let the collection $\{a_0(i), a_1(i),..., a_{l-1}(i)\}$
solves equations of the lattice (\ref{1})
Then by virtue (\ref{compat})
the set of Lax operators
$\{{\cal Q}_i = \hat{w}_i\partial\hat{w}_i^{-1},\; i\in{\bf Z}\}$
admits the action of discrete group ${\cal G}$ with pair of generators
$s_1$ and $s_2$ realized as gauge transformations (\ref{similarity1}) and
(\ref{similarity2}). In addition the group elements
$s_1$ and $s_2$ are restricted by relations
$s_1\circ s_2 = s_2\circ s_1$  and $s_1^q = s_2^p$ where
$p$ and $q$ are co-prime integers such that $pn = qm$, $p\ge 0$.}

Let us now suppose that each ${\cal Q}_{i}$ solves evolution equations
of KP hierarchy (\ref{lax0}). Differentiating the left- and right-hand
sides of auxiliary equations (\ref{aux1}) with respect to $t_p$ by virtue
(\ref{sato}) yields evolution equations
\begin{equation}
\partial_p G_{i} = ({\cal Q}_{i+m}^p)_{+}G_{i} -
G_{i}({\cal Q}_{i}^p)_{+},
\label{eveq-1}
\end{equation}
\begin{equation}
\partial_p H_{i} = ({\cal Q}_{i+n}^p)_{+}H_{i} -
H_{i}({\cal Q}_{i}^p)_{+}.
\label{eveq0}
\end{equation}

Our next goal is to show that the pair of equations
(\ref{eveq-1}) and (\ref{eveq0}) is properly defined and consistent.
To prove correctness of definition of (\ref{eveq0}), standard arguments
are needed  \cite{dickey}. One rewrite (\ref{similarity2}) as
${\cal Q}_{i+n}H_i = H_i{\cal Q}_i$. From this follows
\[
{\cal Q}_{i+n}^pH_i = H_i{\cal Q}_i^{p}
\]
for arbitrary $p\in{\bf N}$. By virtue of this relation one can write
\begin{equation}
({\cal Q}_{i+n}^p)_{+}H_i - H_i({\cal Q}_i^{p})_{+} =
-H_i({\cal Q}_{i+n}^p)_{-} + ({\cal Q}_i^{p})_{-}H_i.
\label{relation}
\end{equation}
The right-hand side of (\ref{relation}) is zero-order $\Psi$DO, while
the left-hand side of (\ref{relation}) is purely differential operator.
So one conclude that the expression
$({\cal Q}_{i+n}^p)_{+}H_i - H_i({\cal Q}_i^{p})_{+}$ is differential operator
of zero-order or simply function.

More complicated situation with (\ref{eveq-1}) since $G_i$'s, generally
speaking, are $\Psi$DO's of special form. However in this situation
one can use equivalent auxiliary system
(\ref{aux5}), with $F_i$ being, as we have mentioned above, purely
differential operator of $l$-order. Evolution equation on $F_i$
follows from (\ref{eveq-1}) and (\ref{eveq0}) and looks as
\begin{equation}
\partial_p F_{i} = ({\cal Q}_{i+m+(l-1)n}^p)_{+}F_{i} -
F_{i}({\cal Q}_{i}^p)_{+}
\label{ev-eq1}
\end{equation}
while the relation
\begin{equation}
{\cal Q}_{i+m+(l-1)n} = F_i{\cal Q}_{i}F_i^{-1}.
\label{cal}
\end{equation}
is valid. Apparently the gauge transformation (\ref{cal}) corresponds
to the group element $s_1\circ s_2^{l-1}\in{\cal G}$. By using the same arguments as
for $H_i$'s one can easily prove that the right-hand side of (\ref{ev-eq1})
is $(l-1)$-order differential operator.

It remains to prove correctness of simultaneous definition of
(\ref{eveq-1}) and (\ref{eveq0}). To do this, it is enough
to show that differentiating the left- and right-hand sides of
(\ref{compat}) with respect to $t_p$, by virtue of
(\ref{eveq-1}) and (\ref{eveq0}) gives identity. It is straightforward
calculation. Let us exhibit some examples of evolution equations
(\ref{eveq-1}) and (\ref{eveq0}) for $t_2$-flows.

{\bf Example 4.1}  Consider the case $n=2, m=1, l=1$ corresponding
to Volterra lattice. $t_2$ flow for operators
$G_i = \partial - r_i$ and $H_i = \partial - r_i - r_{i+1}$ is defined
by evolution equation
\begin{equation}
\partial_2r_{i} = (r_i^{\prime} + r_i^2 + 2r_{i-1}r_i)^{\prime}.
\label{higher}
\end{equation}
By virtue $r_i^{\prime} = r_i(r_{i+1} - r_{i-1})$ from (\ref{higher})
one obtain higher counterpart of Volterra lattice
\[
\partial_2r_{i} = r_i(r_{i+1}r_{i+2} - r_{i-1}r_{i-2} + r_{i+1}^2 - r_{i-1}^2
+ r_ir_{i+1} - r_ir_{i-1}).
\]

{\bf Example 4.2} In the case $n=3, m=1, l=1$ we obtain
\[
\partial_2r_{i} = (r_{i-1}^{\prime} + r_i^{\prime} + r_i^2 +
r_{i-2}r_{i-1} + r_{i-2}r_i + r_{i-1}r_i)^{\prime}
\]
Notice that if we introduce variables $s_i = r_i + r_{i+1}$ then
by virtue
\begin{equation}
r_i^{\prime} + r_{i+1}^{\prime} =
(r_i + r_{i+1})(r_{i+2} - r_{i-1})
\label{virtue}
\end{equation}
the equation
\begin{equation}
\partial_2s_i = s_i(s_{i+2}s_{i+1} - s_{i-1}s_{i-2}).
\label{s_i}
\end{equation}
holds. Notice that (\ref{s_i}) can be obtained as consequence of
\[
\partial_2r_i = s_{i-1}s_is_{i+1} - s_{i-2}s_{i-1}s_i.
\]
By straightforward calculations one can check that $x$- and $t_2$-flows
commute.

Situation in the above example is generalized as follows.
The lattice (\ref{s_i}) come into well known
class of the integrable ones \cite{bogoyavlenskii}
\begin{equation}
\partial_ts_{i} = s_i(\prod_{k=1}^{n-1}s_{i+k} - \prod_{k=1}^{n-1}s_{i-k}),\;
n\ge 2.
\label{chains}
\end{equation}

{\bf Remark.}
It is known that any of the systems (\ref{chains}) can be interpreted
as well as Bogoyavlenskii lattices (\ref{volterra}) as discrete variant of the
Korteweg--de Vries equation \cite{bogoyavlenskii}.

Let $s_i = \sum_{i=1}^{n-1}r_{i+k-1}$. Then one-field
lattices corresponding to the choice $m=1, n\ge 2$ can be written as
$s_i^{\prime} = s_i(r_{i+n-1} - r_{i-1})$.
Notice that (\ref{chains}) is consequence of equation
\[
\partial_tr_{i} =  \prod_{k=0}^{n-1}s_{i-k+1} - \prod_{k=0}^{n-1}s_{i-k}.
\]
Now by straightforward calculations one can check the commutativity of
$x$ and $t$ flows.

Let us show that the systems (\ref{chains}) can be interpreted as restrictions
of 1-Toda lattice flows on corresponding invariant manifolds. In the case
$m=1, n\ge 2, l=1$ eigenvalue problem ${\bf L}(\psi_i) = z\psi_i$ takes on
the form
\begin{equation}
z\psi_{i+n-1} + s_i\psi_{i-1} = z\psi_i.
\label{becomes}
\end{equation}
In the following it is convenient to define new wave functions
by relation $\varphi_i = z^i\psi_i$. In terms of $\varphi_i$'s
auxiliary equation (\ref{becomes}) have following form:
\begin{equation}
\varphi_{i+n-1} + s_iz^{n-1}\varphi_{i-1} = z^{n-1}\varphi_i.
\label{form}
\end{equation}
Step-by-step one can expand the left-hand side of equation (\ref{form})
to obtain eigenvalue problem ${\bf M}(\varphi_i) = z^{n-1}\varphi_i$, where
\begin{equation}
{\bf M} = E^{n-1} +
\sum_{j=1}^{\infty}\left(\prod_{k=1}^js_{i-j+1}\right)E^{n-j-1}
\label{where}
\end{equation}
with $E$ being an operator of elementary shift $E(\eta_i) = \eta_{i+1}$.

For each $n\ge 2$ define ${\bf A}_n = {\bf M}_{+}$, where subscript $+$
stands for projection on positive part of ${\bf M}$. The difference
operator ${\bf A}_n$ is used to define auxiliary evolution equation
$\partial\varphi_i/\partial t = {\bf A}_n(\varphi_i)$. It is easy
to verify that the latter is consistent with (\ref{form}) provided
that equations (\ref{chains}) satisfy. From this one can conclude
that the Lax equation $\partial_t{\bf M} = [{\bf A}_n, {\bf M}]$
is equivalent to the system (\ref{chains}).

Let us explain how above is relevant to 1-Toda lattice hierarchy \cite{ueno}.
For Lax operator ${\bf Q} = E + \sum_{k\ge 0}q_k(i)E^{-k}$ one defines
restriction ${\bf Q}^{n-1} = {\bf M}$, where ${\bf M}$ is in
(\ref{where}). This implies that some algebraic constraints
on coefficients $q_k(i)$ must be imposed. For example in the simplest
case of Volterra lattice ($n=2$) these constraints are given by
\[
q_k(i) = s_{i-k}...s_{i-1}s_i =
q_0(i-k)...q_0(i-1)q_0(i),\;\; k\ge 1.
\]

To conclude this section, let us consider one example which illustrate
relationship between a class of the lattices (\ref{1}) and other known
lattices. In the case $n=1, m=-1, l=3$ the equations (\ref{1}) take on the form
\begin{equation}
\begin{array}{l}
a_0^{\prime}(i) + a_0^{\prime}(i+1) = a_0^2(i) - a_0^2(i+1) + a_1(i+1) - a_1(i), \\[0.4cm]
a_1^{\prime}(i) = a_2(i+1) - a_2(i), \\[0.4cm]
a_2^{\prime}(i) = a_2(i)(a_0(i) - a_0(i-1))
\end{array}
\label{MB}
\end{equation}
Using (\ref{identify}) one calculates Miura transformation between
(\ref{MB}) and (\ref{virtue}) to obtain
\begin{equation}
\begin{array}{l}
a_0(i) = r_{3i} + r_{3i+1} + r_{3i+2}, \\[0.4cm]
a_1(i) = (r_{3i-1} + r_{3i})(r_{3i} + r_{3i+1}) \\[0.4cm]
+ (r_{3i} + r_{3i+1})(r_{3i+1} + r_{3i+2}) +
(r_{3i+1} + r_{3i+2})(r_{3i+2} + r_{3i+3}), \\[0.4cm]
a_2(i) = (r_{3i-2} + r_{3i-1})(r_{3i-1} + r_{3i})(r_{3i} + r_{3i+1}).
\end{array}
\label{MBBT}
\end{equation}
Higher counterpart of (\ref{MB}) is nothing but
Blaszak--Marciniak lattice \cite{blaszak}
\begin{equation}
\begin{array}{l}
\partial_2p_i = u_{i+2} - u_i, \\[0.4cm]
\partial_2v_i = p_iu_{i+1} - u_ip_{i-1}, \\[0.4cm]
\partial_2u_i = u_i(v_i - v_{i-1}),
\end{array}
\label{MBB}
\end{equation}
with $\partial_2a_0(i) = u_{i+1} - u_i$.
Here we denote $p_i = a_0(i) + a_0(i+1), v_i = a_1(i), u_i = a_2(i)$.
>From (\ref{MBBT}) we easy obtain Miura transformation
\[
\begin{array}{l}
p_i = s_{3i} + s_{3i+2} + s_{3i+4}, \\[0.4cm]
v_i = s_{3i-1}s_{3i} + s_{3i}s_{3i+1} + s_{3i+1}s_{3i+2}, \\[0.4cm]
u_i = s_{3i-2}s_{3i-1}s_{3i}.
\end{array}
\]
between (\ref{MBB}) and (\ref{s_i}).

\section{Conclusion}

Starting from auxiliary linear equations (\ref{aux2}), we have defined
a class of integrable first-order $l$-field lattices. The main
feature shared by the latter is in some sense compatibility with KP
flows. Taking this fact into account, one can exploit natural idea to search
for solutions to the above lattices and its higher counterparts in terms
of the KP $\tau$-functions.
In future we are going to continue our activity in this direction.
It will be of interest to investigate such a questions as lattice
B\"acklund transformations and nonlinear superposition formulas.

The results of the paper might be of potential interest for
investigation of constrained KP hierarchies. We believe that
any integrable chains (up to Miura transformations)
presented in the paper underlie some differential integrable
hierarchies with $s_1$ and $s_2$ being discrete symmetries.
Activity in this direction is in the paper \cite{svinin1}
(see also \cite{svinin2}) where we have constructed modified version
of Krichever's rational reductions of KP hierarchy. Moreover,
the discrete symmetries $s_1$ and $s_2$ in this case as was shown
in \cite{svinin1} correspond to one-field lattices with $m\in{\bf N}$.

\section*{Acknowledgments}
We are grateful to anonymous referee for carefully reading the manuscript
and for remarks which enabled us to improve the presentation of the paper.
This research has been supported by INTAS grant 2000-15.

The author
wish to thank Editorial Board for invitation to contribute the paper
to this issue.

\end{document}